\documentstyle[12pt]{article}

\begin{document}

{\flushright{\small Imperial/TP/97-98/54\\
    hep-th/9806089\\}}

\begin{center}
  {\Large {\bf The central charge in three dimensional anti-de Sitter space}}
  
  \vspace{0.5in}
  
  Maximo Ba\~nados\footnote{On leave from  Centro de Estudios
    Cient\'{\i}ficos de Santiago, Casilla 16443, Santiago, Chile and,
    Departamento de F\'{\i}sica, Universidad de Santiago, Casilla 307,
    Santiago, Chile.}  \vskip 0.2cm {{\it Departamento de F\'{\i}sica
      Te\'orica, Facultad de Ciencias,\\ Universidad de Zaragoza,
      Zaragoza 50009, Spain}
    \\
    \small E-mail: {\tt max@posta.unizar.es}} \vspace{0.2in}
  
  Miguel E. Ortiz \vskip 0.2cm {{\it Blackett Laboratory, Imperial
      College of Science, Technology
      and Medicine,\\
      Prince Consort Road, London SW7 2BZ, UK}
    \\
    \small E-mail: {\tt m.ortiz@ic.ac.uk}} \vskip 0.5truecm %\today
\end{center}

\vskip 0.5truecm

\begin{abstract}
 This paper collects the various ways of computing the central
  charge $c=3l/2G$ arising in 3d asymptotically anti-de Sitter spaces,
  in the Chern-Simons formulation. Their similarities and
    differences are displayed.
\end{abstract}

\medskip

\newpage

The recently conjectured duality \cite{Maldacena} between anti-de
Sitter space and conformal field theory has opened the door to many
surprising connections between those two theories. An adS/CFT
correspondence was first obtained by Brown and Henneaux \cite{BH} who
proved that the asymptotic symmetry group of anti-de Sitter space in
three dimensions is the conformal group with a classical central
charge 
\begin{equation}
c=\frac{3l}{2G}.
\label{c}
\end{equation}
This observation led Strominger \cite{Strominger} (see also
\cite{Birmingham}) to consider the degeneracy of states with $L_0$ and
$\bar L_0$ fixed and he found the surprising result that its logarithm
agrees exactly with the Bekenstein-Hawking entropy of a three
dimensional asymptotically anti-de Sitter black hole \cite{BTZ}. The
connection between CFT and black hole entropy can be seen by looking
at the semiclassical form of the free energy $f$ of a non-rotating
three dimensional black hole with inverse temperature $\beta$ (in
units of $l$),
\begin{equation}
-\beta f = -\beta e + \frac{ \pi c }{6 \beta} 
\end{equation}
where $e$ is proportional to the black hole mass and $c$ is given by
(\ref{c}). This coincides with the free energy of a strip of width
$\beta$ (with periodic boundary conditions) in a 1+1 CFT \cite{Cardy}
with central charge $c$.

The Virasoro algebra discovered in Ref. \cite{BH} arose by studying
the asymptotic Killing equations for anti-de Sitter spacetimes. It was
then realized \cite{B} that using the Chern-Simons formulation of 2+1
gravity and its connection to WZW theories, a simple realization of
the conformal generators (by means of a twisted Sugawara construction)
was possible. The central charge is associated in \cite{B} to
diffeomorphisms with a component normal to the boundary. The
underlying boundary theory in this approach is a chiral WZW model
having a Kac-Moody symmetry. The Virasoro algebra is then a
sub-algebra of the Kac-Moody enveloping algebra.  Since in this space
there is a family of Virasoro operators with different classical
central charges, in order to fix $c$, or equivalently, to fix the
relevant twisted Sugawara operator more information is needed. Fixing
the Virasoro algebra to leave the asymptotic form of anti-de Sitter
space invariant fixes $c=6k$ \cite{BBO}, where $k$ is the level of the
Chern-Simons theory.  Since the equivalence between the gravitational
and Chern-Simons actions fixes $k=l/4G$, the central charge is then
the same as that calculated in \cite{BH}.  We shall come back to this
point below.

A different Chern-Simons formulation for the results of \cite{BH} was
proposed in \cite{CHvD} by making use of the known reduction from
$WZW_{SL(2,\Re)}$ theory (obtained at the boundary from the
Chern-Simons theory) to Liouville theory \cite{Forgacs-}.  Liouville
theory has two commuting Virasoro operators with a classical central
charge equal to $c=6k$, where $k$ is again the level of the
Chern-Simons theory, and the central charge is therefore again the
same as that calculated in \cite{BH}. It is remarkable that the
conditions that reduce $WZW_{SL(2,\Re)}$ to Liouville theory are
indeed satisfied by anti-de Sitter space asymptotically. In this
approach the underlying Hilbert space then corresponds (at least
ignoring zero modes) to that of a Liouville field and the central
charge is uniquely determined to be $c=6k$.  (This approach has
recently been extended to supergravity in \cite{BBCHO}.)

The existence of a Kac-Moody algebra at the boundary was thought to be
useful in giving a microscopical derivation of the entropy of three
dimensional black holes \cite{Carlip} (see also \cite{B,BG,BBO})
independently of the connection with the recent work of Strominger,
and provides a natural arena to explore 2+1 quantum gravity. It is
surprising that Carlip's approach to computing the black hole's
entropy does not depend so much on the value of $c$ as on the
Kac-Moody structure. Conversely, Strominger's idea \cite{Strominger}
depends crucially on the value of $c$, while the underlying Kac-Moody
symmetry seems to be unimportant, and from this point of view can even
be interpreted as predicting a state counting that disagrees with the
Bekenstein-Hawking result \cite{BBO,Car}. For this reason, we find it
useful to have a clear relation between the boundary theories found in
\cite{B} and \cite{CHvD}.

This letter is a collection of different known results including
elements of \cite{Forgacs-,Polyakov} (and the work that followed those
papers) applied to 2+1 dimensional gravity and its relation to chiral
WZW models.  Our main goal is to compare and exhibit the similarities
and differences between the approaches followed in \cite{B} and
\cite{CHvD} to describe the conformal generators and the central
charge discovered in \cite{BH}.

Three dimensional adS gravity can be recast as a $SL(2,\Re)\times
SL(2,\Re)$ Chern-Simons theory \cite{Achucarro-}.  Consider first one
$SL(2,\Re)$ \cite{not} Chern-Simons theory formulated on a manifold with the
topology of a solid cylinder. We use polar coordinates
$\{r,t,\varphi\}$ on the cylinder. The boundary is located at
$r\rightarrow\infty$ and has the topology of a cylinder. On the
cylinder we use coordinates $x^\pm = \pm t + \varphi$. The
Chern-Simons action needs boundary conditions on the cylinder to be
well defined. We take these boundary conditions from anti-de Sitter
space. Firstly we require that the radial component of the gauge field
be given by
\begin{equation}
A_r = \gamma H
\label{Ar}
\end{equation}
where $H$ is the generator of the Cartan sub-algebra and $\gamma$ is a
real constant. This condition follows directly from the gauge field
representing anti-de Sitter space.  Note that this fixes
$(A_r)^2=\gamma^2/2$ (in the notation of \cite{B,BBO}, this is
$\alpha^2$) and that $\gamma$ parametrises a rescaling of the radial
coordinate at infinity. Although we could fix $\gamma$ by working with
proper radial coordinates at infinity \cite{BBO} (the relation between
this radial coordinate and the one used in \cite{CHvD} is $r=\log
\tilde r$), we keep an explicit dependence on $\gamma$ to emphasise
the fact that the final central charge is independent of it. Secondly,
we impose the condition
\begin{equation}
A^a_-=0
\label{A-}
\end{equation}
which also follows from classical adS space. (The corresponding
boundary condition for the other $SL(2,\Re)$ copy is $\tilde A_+=0$.)

It is well known that these boundary conditions lead to a Kac-Moody
algebra for the currents $A_\varphi(x^+)$.  The group of gauge
transformations leaving (\ref{Ar}) and (\ref{A-}) invariant are
characterised by parameters $\lambda(x^+,r)= \exp(-r H) \lambda(x^+)
\exp(r H)$ (they do not depend on $x^-$ and the radial dependence can
be gauged away), and are generated by Kac-Moody currents. In the
following we shall denote the current $A_\varphi$ simply by $A$.

The point raised in \cite{B} is that by considering the subset of
gauge transformations with parameters
\begin{equation}
\lambda^a = -f' H^a + f A^a = -{f'\over \gamma}A^a_r + f A^a
\label{1}
\end{equation}
with $f$ an arbitrary function of $x^+$ (prime denotes derivative),
the associated algebra is the Virasoro algebra with $c = 6k$. 
  (Note that the way this formula was presented in \cite{B,BBO} might
  have given the impression that the value of $c$ depends on $\gamma$
  but this was because the formula given there assumed the use of a
  proper radial coordinate at infinity.)  This can be seen simply by
noticing that the associated generator is a twisted Sugawara operator
\cite{B}
\begin{equation}
L =  \mbox{Tr} \left(\frac{1}{2} A^2 + H A' \right) 
\label{L}
\end{equation}
with $A^a$ being the $SL(2,\Re)$ Kac-Moody currents. This is not a
coincidence.  The parameters of the form (\ref{1}) can be interpreted
as diffeomorphisms via the relation $\lambda^a = \xi^i A^a_i$.
Comparing with (\ref{1}) we find $\xi^i = \{-f'/\gamma,f \}$ which was
shown in \cite{BBO} to be exactly the form of the residual
diffeomorphisms found in \cite{BH} (up to a radial
reparameterisation).  We thus arrive at the
conclusion that the residual group of diffeomorphisms (asymptotic
Killing vectors) found in \cite{BH} has a natural action in the
connection representation through a twisted Sugawara construction.

However, as stressed above, the symmetry algebra leaving invariant the
boundary conditions (\ref{Ar}) and (\ref{A-}) is the whole Kac-Moody
symmetry within which (\ref{L}) only generates a sub-algebra of the
enveloping algebra which explicitly leaves invariant the
  asymptotic form of the metric. There is in fact a whole family of
Virasoro generators with different central charges which can be
obtained by multiplying the first term in (\ref{1}) by a constant. The
particular form of (\ref{L}) is selected because it generates
diffeomorphisms that coincide with the asymptotic symmetries found in
\cite{BH}.  The value of the central charge, $c=6k$, is thus
associated with the invariance properties of the asymptotic metric. In
this sense, this procedure can be viewed as re-deriving \cite{BH} from
the connection point of view (see \cite{BBO} for a detailed discussion
on this point), rather than as using any intrinsic property relating
the Kac-Moody and Virasoro algebras.

In \cite{CHvD} a different route was taken to relating the Kac-Moody
and Virasoro algebras.  First it was proved that the two $SL(2,R)$
chiral WZW models (arising in 2+1 dimensional gravity with negative
cosmological constant) can be glued together into a single non-chiral
one.  It then follows that imposing the extra conditions \cite{Forgacs-}
\begin{equation}
A^-=1, \ \ \  \tilde A^+=1,  
\label{2'}
\end{equation}
plus $A^H=0=\tilde A^H$ leads to a Liouville theory (up to zero modes
contributions that have not been calculated so far, and using the
Gauss decomposition to parametrize the group manifold).  It is
remarkable that the conditions (\ref{2'}) are satisfied asymptotically
by the anti-de Sitter gauge field \cite{CHvD}.  Liouville theory is
conformally invariant and it has two commuting Virasoro operators with
a classical central charge $c=6k$ \cite{Forgacs-}. This procedure thus
fixes the value of $c$ with no ambiguity, although of course the use
of (\ref{2'}) in the present context depends on its remarkable
relation to the asymptotics of adS$_3$.

Another derivation of the two Virasoro operators with $c=6k$,
without resorting to Liouville theory, is obtained by working directly
with each chiral $SL(2,\Re)$ WZW model. As shown in \cite{Polyakov}  
the conditions 
\begin{equation} A^-=1, \ \ \ \ A^H=0,
\label{2}
\end{equation}
transmute an affine $SL(2,\Re)_k$ algebra into a Virasoro algebra with
$c=6k$.  This can be seen as follows \cite{Polyakov}. The parameters
of the most general set of gauge transformations leaving (\ref{2})
invariant have the form (in the basis $\{J_\pm,H\}$)
\begin{equation} \lambda = \left(  \begin{array}{cl}
                        (1/2)f' &  f A^+ - (1/2) f'' \\
                                                f  &  -(1/2)f'
            \end{array} \right)
\label{f}
\end{equation}
where $f$ is an arbitrary function of $x^+$. It is straightforward to
check (using Dirac brackets or otherwise) that this subset of gauge
transformations acting on the remaining component $A^+$ yields a
Virasoro algebra with $c=6k$ \cite{Polyakov} (see \cite{Bais-} for a
generalization). An advantage of this procedure is that no coordinates
on the group manifold have been used. The Virasoro generator $A^+$ is
thus well adapted to deal with global issues, although no action has
been written for it. Details of this procedure as applied to the
supergravity/superCFT correspondence can be found in Ref.
\cite{BBCHO}.

Our last step is to prove that once (\ref{2}) is imposed, the twisted
Virasoro operator (\ref{L}) reduces to $A^+$, and that the parameters
(\ref{1}) reduce to (\ref{f}). In other words, on the constraint
surface defined by (\ref{2}), the quadratic Virasoro operator
(\ref{L}) is equivalent to $A^+$ and, of course, they satisfy the same
algebra. First we need to study the transformations $\delta A^a = D
\lambda^a$ with $\lambda^a$ given by (\ref{1}) and see whether they
are consistent with the restrictions (\ref{2}).  If they are not
consistent, then we need to modify the transformation (\ref{1}). This
is analogous to the reduction from Kac-Moody to Virasoro via
(\ref{2}). One cannot just plug (\ref{2}) into the Kac-Moody algebra
and expect to get the Virasoro algebra for the remaining component
$A^+$.  The right thing to do it is to construct the Dirac bracket
whose dynamics is consistent with the constraints. Then, one finds the
Virasoro algebra (in the Dirac bracket) with a non-zero central
charge.

The variation of $A^a$ with respect to the gauge transformation
(\ref{1}) in components is
\begin{eqnarray}
\delta A^+ &=& (f A^+)' + A^+ f' \label{+} \\
\delta A^- &=& f (A^-)' \label{-} \\
\delta A^3 &=& (fA^3)' - f'' \label{3}
\end{eqnarray}
We see that the condition $A^-=1$ causes no problems since it is
preserved by (\ref{-}) (this can also be seen from the fact that $A^-$
commutes (weakly) with the modified Sugawara operator).  On the other
hand $A^3=0$ is not preserved by (\ref{3}) due to the $f''$ term.

The analogue of constructing the Dirac bracket in this calculation is
to modify the parameter (\ref{2}). Consider gauge transformations with
an improved parameter
\begin{equation}
\lambda_i = -f' H + f  A - \frac{1}{2} f'' J_+.
\label{i}
\end{equation}
This transformation is a combination of a diffeomorphism and a gauge
transformation \cite{BBO}.  Studying the variation of the connection
under these improved transformations, one finds
\begin{eqnarray}
\delta_i A^+ &=& (f A^+)' + A^+ f' - (f''/2) A^3 - (1/2) f'''\label{i+} \\
\delta_i A^- &=& f (A^-)'  \label{i-} \\
\delta_i A^3 &=& (fA^3)' + f''(-1+A^-) \label{i3}
\end{eqnarray}
We see that the improved transformations are consistent with the
constraints (\ref{2}) and the residual function $A^+$ transforms as a
Virasoro generator with a classical central term $c=6k$. Now, we can
see that after imposing (\ref{2}) the parameter (\ref{i}) reduces
exactly to the residual symmetries (\ref{f}).  Similarly, the Sugawara
operator (\ref{L}) reduces directly to
$A^+$, as expected. \\

We are grateful to Marc Henneaux for many informative and stimulating
discussions on anti-de Sitter asymptotics, and for a critical reading
of the manuscript. We also thank T. Brotz and S. Carlip for useful
conversations.


\begin{thebibliography}{999}

\bibitem{Maldacena} J. Maldacena, {\em The large N limit of superconformal
field theories and supergravity}. 
hep-th/9711200.

\bibitem{BH} J.D. Brown and M. Henneaux, {\em Commun.
Math. Phys.} {\bf 104}, 207 (1986).

\bibitem{Strominger} A. Strominger, {\em Black hole entropy from
    near-horizon microstates}. hep-th/9712251.  
        
\bibitem{Birmingham}
  D. Birmingham, I. Sachs and A. Sen, {\em Entropy of
    three-dimensional black holes in string theory}. hep-th/9801019.    

\bibitem{BTZ} M. Ba\~nados, C. Teitelboim and J.Zanelli,
              Phys. Rev. Lett. {\bf 69}, 1849 (1992).
                          
\bibitem{Cardy} H.W. Blote, J.L. Cardy and M.P. Nightingale, 
Phys. Rev. Lett. {\bf 56}, 742 (1986).  I. Affleck, Phys. Rev. Lett. {\bf 56}, 746 (1986).        

\bibitem{B} M. Ba\~nados,  Phys. Rev. {\bf D52}, 5816 (1995); hep-th/9405171.

\bibitem{BBO} M. Ba\~nados, T. Brotz and M. Ortiz, {\em Boundary 
dynamics and the statistical mechanics of the 2+1 dimensional 
black hole}. hep-th/9802076

\bibitem{CHvD} O. Coussaert, M. Henneaux, P. van Driel 
{\em Class. Quant. Grav. } {\bf 12}, 2961 (1995); gr-qc/9506019. 

\bibitem{Forgacs-} P.  Forg\'acs, A. Wipf, J. Balog, 
L. Feh\'er and L. O'Raifeartaigh, {\em
      Phys. Lett.} {\bf 227 B} (1989) 214.

\bibitem{Achucarro-} A. Ach\'ucarro and 
P.K. Townsend, {\em Phys. Lett.} {\bf B180}, 89 (1986).

\bibitem{not} We use the $SL(2,\Re)$ basis $\{J_\pm,H\}$
  satisfying $[J_+,J_-]=2H$, $[H,J_\pm]=\pm J_\pm$ and Tr$H^2=1/2$.
  The $SL(2,\Re)$ gauge field $A$ is decomposed as $A=A^+J_+ + A^-J_-
  + A^H H$.

\bibitem{BBCHO} M\'aximo Ba\~nados, Karin Bautier, Olivier 
Coussaert, Marc Henneaux and Miguel Ortiz, 
{\em Anti-de Sitter/CFT correspondence in three dimensional supergravity}.
hep-th/9805165

\bibitem{Carlip}  S. Carlip, Phys. Rev. {\bf D51}, 632 (1995);  
                  S. Carlip, Phys. Rev. {\bf D55}, 878 (1997) 

\bibitem{BG}  M. Ba\~nados and A. Gomberoff, 
              Phys. Rev. {\bf D55}, 6162, (1997)
              
\bibitem{Car} S. Carlip, {\em What we dont know about BTZ
                black hole entropy}. hep-th/9806026.

\bibitem{Polyakov} A.M. Polyakov,
  {\em Int. J. Mod. Phys.} {\bf 5} (1990) 833.

\bibitem{Bais-} F.A. Bais, T.Tjin and P. van Driel, Nucl. Phys. {\bf B357}, 632 (1991). 


\end{thebibliography}
\end{document}